\documentclass[aps,pr,twocolumn,groupedaddress,notitlepage,%showpacs,
floatfix,superscriptaddress]{revtex4-1}

\pdfoutput=1
\usepackage{graphicx,graphics,epsfig,subfigure,times,bm,bbm,amssymb,amsmath,amsthm,mathrsfs,MnSymbol}
\usepackage{gensymb}
\usepackage{amsfonts}
\usepackage{float}
\usepackage[matrix,frame,arrow]{xypic}
\usepackage[pdfstartview=FitH]{hyperref}

\usepackage[pdftex]{color}
\usepackage{braket}  %Dirac Notation in QM
\usepackage{enumerate}
\usepackage[normalem]{ulem}
\usepackage[usenames,dvipsnames]{xcolor}
\usepackage{multirow}
\usepackage{mathtools}
\usepackage{bbm}
\usepackage{titletoc}

\definecolor{orange}{rgb}{1,0.5,0}
\newcommand{\RNum}[1]{\uppercase\expandafter{\romannumeral #1\relax}}

\newcommand{\ignore}[1]{}
\usepackage{geometry}

\ignore{
	\documentclass[eprintnumbers,amsmath,amssymb,onecolumn,a4paper,caption ]{article}
	\usepackage{amsfonts}
	\usepackage{amssymb}
	\usepackage{mathrsfs}
	\usepackage{mathbbold}
	\usepackage{bbm}
	\usepackage{mathrsfs}
	\usepackage{dcolumn}% Align table columns on decimal point
	\usepackage{bm}% bold math
	\usepackage{times,epsfig,amssymb,amsmath}
	\usepackage{float}
	\usepackage{subfigure}
	\usepackage{geometry}\geometry{left=2.5cm,right=2.5cm,top=3cm,bottom=3cm}

	\usepackage{color}

}

\hypersetup{
	colorlinks=true,       % false: boxed links; true: colored links
	linkcolor=red,          % color of internal links
	citecolor=blue,        % color of links to bibliography
	filecolor=magenta,      % color of file links
	urlcolor=blue,           % color of external links
	runcolor=cyan
}
\begin{document}
	\title{Robustness and Independence of the Eigenstates with respect to the Boundary Conditions across a Delocalization-Localization Phase Transition}

	\author{Zi-Yong~Ge}
	\affiliation{Beijing National Laboratory for Condensed Matter Physics, Institute of Physics, Chinese Academy of Sciences, Beijing 100190, China}
	\affiliation{School of Physical Sciences, University of Chinese Academy of Sciences, Beijing 100190, China}

	\author{Heng~Fan}
	\email{hfan@iphy.ac.cn}
	\affiliation{Beijing National Laboratory for Condensed Matter Physics, Institute of Physics, Chinese Academy of Sciences, Beijing 100190, China}
	\affiliation{School of Physical Sciences, University of Chinese Academy of Sciences, Beijing 100190, China}
	\affiliation{CAS Center for Excellence in Topological Quantum Computation, UCAS, Beijing 100190, China}
	\affiliation{Songshan Lake Materials Laboratory, Dongguan 523808, China}

	\begin{abstract}
	We focus on the many-body eigenstates across a localization-delocalization phase transition.
	To characterize the robustness of the eigenstates, we introduce eigenstate overlaps $\mathcal{O}$ with respect to different boundary conditions.
	In the ergodic phase, the average of eigenstate overlaps $\bar{\mathcal{O}}$ 
	exponentially decays with an increase in the system size, indicating the fragility of its eigenstates,
	and this can be considered as an eigenstate version of the butterfly effect of chaotic systems.
	For localized systems, $\bar{\mathcal{O}}$ is almost size independent, 
	showing the strong robustness of the eigenstates and the inconsistency of the eigenstate thermalization hypothesis.
	In addition, we find that the response of eigenstates to the change in the boundary conditions  in 
	many-body localized systems is identified with the single-particle wave functions in Anderson localized systems.
	This indicates that the eigenstates of many-body localized systems, as  many-body wave functions, may be independent of each other.
	We demonstrate that this is consistent with the existence of a large number of quasilocal integrals of motion in the many-body localized phase.
	Our results provide another method to study localized and delocalized systems from the perspective of eigenstates.

	\end{abstract}

	\date{\today}
	\maketitle

    \section{Introduction}
    Recently, understanding the mechanisms of thermalization and localization in isolated quantum many-body 
    systems has attracted much interest.
    Generally, for a closed quantum many-body system, starting from a far-from-equilibrium initial state, 
    the system can always thermally equilibrate under a unitary evolution~\cite{PhysRevA.43.2046,PhysRevE.50.888,Rigol2008Nature,DAlessio2016,Gogolin_2016}.
    We call these ergodic systems, and the microscopic mechanism of this thermalization is known as the eigenstate thermalization hypothesis (ETH).
    Nevertheless, there also exist localized systems, which are typical examples violating ETH~\cite{Rahul2015}.
    The localized systems were first identified by Anderson in a non-interacting fermion system with impurity scatterings, which was dubbed Anderson localization (AL)~\cite{PhysRev.109.1492}.
    In the past two decades, it was shown that the localization can persist in the presence of interactions, which is now termed many-body localization (MBL)~\cite{Basko2006,PhysRevB.75.155111,PhysRevLett.95.206603,PhysRevB.82.174411,PhysRevB.90.174202,Ehud2018NatPhy,RevModPhys.91.021001}.
    Benefiting from experimental advances in synthetic quantum many-body systems,
     MBL has been realized in various platforms, such as optical lattices~\cite{Schreiber842,Choi1547,PhysRevLett.116.140401}, 
    nuclear magnetic resonance~\cite{Alvarez846}, trapped ions~\cite{JSmith2016}, and superconducting circuits~\cite{PhysRevLett.120.050507,arxiv1912.02818}.

    Comparing with ergodic and AL systems, MBL systems possess many unique properties.
    For the dynamics, the MBL system can hardly be thermalized 
    due to the existence of a large number of quasilocal integrals of motion~\cite{PhysRevLett.111.127201,PhysRevLett.116.010404},
    and the entanglement entropies can exhibit long-time logarithmic spreading~\cite{PhysRevLett.109.017202,PhysRevLett.110.260601}, while the growth of entanglement is ballistic in ergodic systems~\cite{PhysRevLett.111.127205}.
    Additionally, according to the level statistics, it is shown that the spectrum of MBL systems obeys a Poisson distribution,  
    while it is a Wigner-Dyson distribution in the ergodic phase~\cite{RevModPhys.69.731,PhysRevB.75.155111,PhysRevB.82.174411,PhysRevB.90.064203,PhysRevLett.122.180601}.
    Furthermore, the eigenstates of MBL systems also have many unique properties, especially the entanglement. It is shown that the eigenstates of the MBL systems have low entanglement, where the entanglement entropies satisfy the area law, and the entanglement spectrum is a power law~\cite{PhysRevLett.111.127201,PhysRevB.90.174202,Imbrie2016,PhysRevLett.117.160601}.

    In Ref.~\cite{Edwards_1972}, the authors use the sensitivity of the single-particle eigenenergies to choose periodic or antiperiodic boundary conditions as a criterion to identify the AL phase.
    In addition, to further study the  sensitivity of the spectrum toward small perturbations, the concept of \textit{level curvatures} was introduced~\cite{PhysRevE.47.1650,PhysRevLett.73.798,fyodorov1995universality,Titov_1997,PhysRevE.99.050102,PhysRevB.99.224202}, which can  be applied to delocalized and localized systems.
    In Ref.~\cite{PhysRevX.5.041047}, the authors find that the distributions of the off-diagonal matrix elements of a local operator 
    are distinct between ergodic and MBL systems, and it can be a criterion to probe the MBL phase transition.   
    There are also a few works to discuss the behaviors of eigenstates under small perturbations.
    For instance, in Ref.~\cite{PhysRevE.86.046202}, the authors study the \textit{overlap matrix} and \textit{overlap distance} of the random matrix model, which can be used for comparing the eigenspaces of perturbed and nonperturbed Hamiltonians. The entries of the \textit{overlap matrix} are obtained by calculating the overlaps of two arbitrary  eigenstates of a perturbed and nonperturbed Hamiltonian, respectively. 
    In Ref.~\cite{Monthus_2017}, the concept of \textit{fidelity susceptibility} was introduced, which is the second derivative of the diagonal entries of the \textit{overlap matrix} with respective to the strength of the perturbations, and this object can be used to probe the MBL.
    However, it is still an open question whether and how the responses of the eigenstates towards the change in boundary conditions
    can distinguish delocalized, AL, or MBL systems.
    
    	 \begin{figure}[t]   
    	\includegraphics[width=0.45\textwidth]{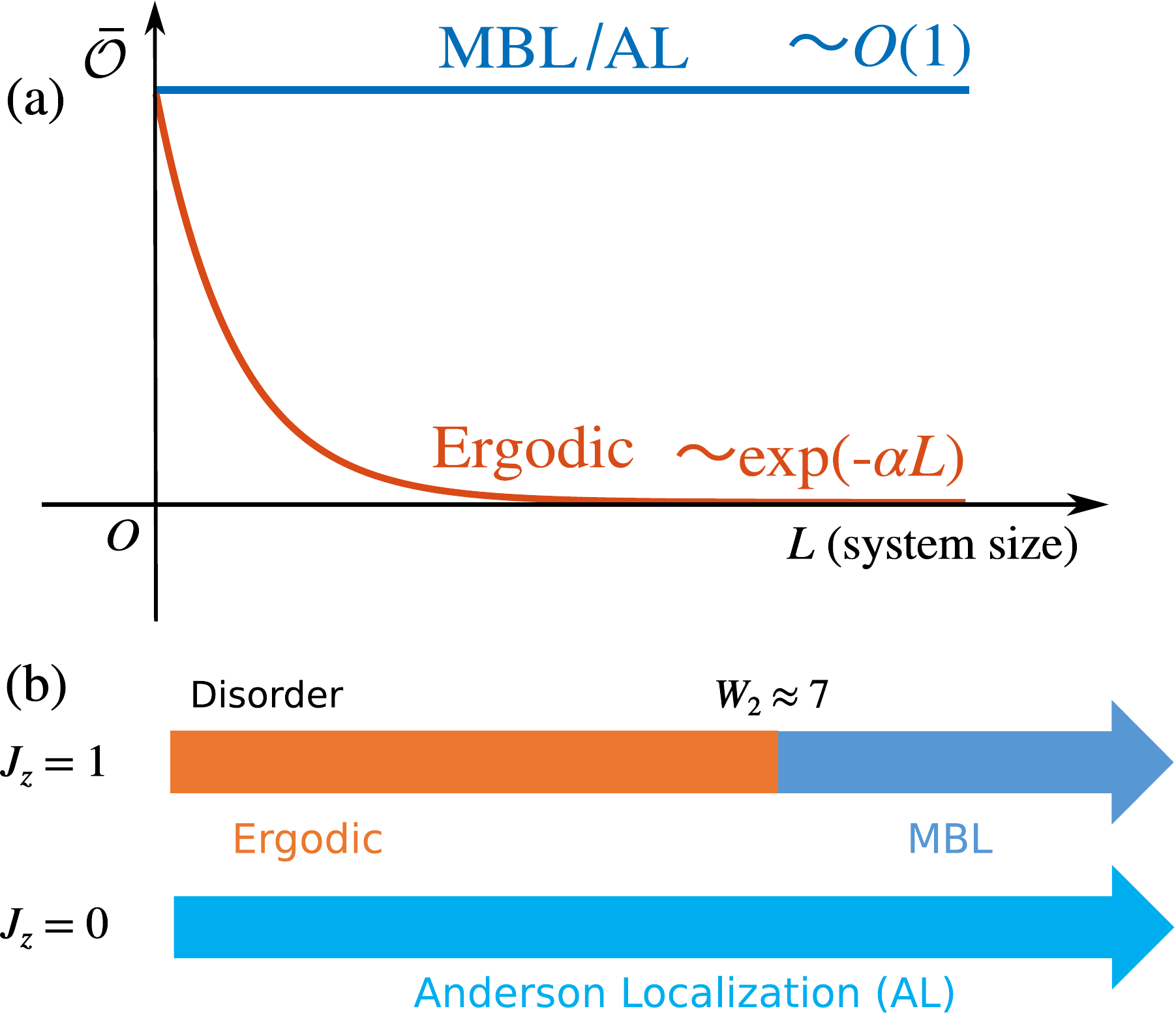}
    	\caption{(a) The scaling functions of $\bar{\mathcal{O}}$ for ergodic, AL, and MBL phases, respectively.
    		(b) The phase diagram of Hamiltonian (\ref{H}). Here, the Hamiltonian is defined by Pauli matrices rather than spin-$\frac{1}{2}$ operators, so the disorder strength $W$ is twice as large relative to  Ref.~~\cite{PhysRevB.82.174411}. }
    	\label{fig_1}
    \end{figure}

    In this paper, we investigate the responses of eigenstates
    with respective to the change of boundary conditions in the localized and delocalized systems.
    We mainly calculate the overlaps of the corresponding eigenstates between two different boundary conditions.
    We find that the eigenstates of the MBL and AL systems show strong robustness, since the eigenstate overlaps are nearly size independent. 
    Nevertheless, in the ergodic phase, the eigenstate overlaps decay exponentially with an increase in system size, which can be considered as a butterfly effect of quantum chaos. These results are summarized in Fig.~\ref{fig_1}(a).
    In addition, we find that the responses of many-particle eigenstates to the change in the boundary conditions in MBL systems are akin to the one of single-particle eigenstates in AL systems.
    Thus, we suppose that the  eigenstates of MBL systems are independent of each other, which is distinct from AL systems, whose many-particle eigenstates are Slater determinants.
    We also demonstrate that this is consistent with the existence of a large number of quasi-local integrals of motion.

    The remainder of this paper is organized as follows: In Sec.~\ref{Model}, we introduce a one-dimensional spin-$\frac{1}{2}$ XXZ model with a $z$-directed random field, and the corresponding phase diagrams are reviewed. We also present the main methods of this paper.
    The numerical results are shown in Sec.~\ref{Numerical results}. 
    In Sec.~\ref{Phenomenological discussion}, we give a phenomenological discussion about our results.
    Finally, in Sec~\ref{Conclusion}, we summarize our results and present an outlook for future research.
    Additional numerical results are present in the Appendix.

	\section{Model and Methods}\label{Model}
	In this section, we introduce the random-field XXZ chain, the main model studied in this paper, and then provide the numerical methods.
	For our methods, the eigenstate overlaps $\mathcal{O}$ with respect to the different boundary conditions are defined, and this can be considered as the measurement of the response of the eigenstates with respect to the change in the boundary conditions.

   \subsection{Model}	
	 We consider a spin-$\frac{1}{2}$ XXZ chain with a $z$-directed random field. 
	 The Hamiltonian of this model reads
	 	\begin{align} \label{H} \nonumber
	 	\hat H= &J\sum_{i}^{L} (\hat{\sigma}^x_{i} \hat{\sigma}^x_{i+1}+\hat{\sigma}^y_{i}
	 	\hat{\sigma}^y_{i+1})+ J_z\sum_{i}^{L-1}\hat{\sigma}^z_{i} \hat{\sigma}^z_{i+1}\\
	 	&+\sum_{i=1}^Lh_{i}\hat{\sigma}^z_{i},
	 	\end{align}
	 where $\hat{\sigma}^\alpha$'s ($\alpha=x,y,z$) are Pauli matrices.
	 The random field $h_i \in [-W,W]$ satisfies a uniform distribution.
	 The localization-delocalization transition of this model has been studied extensively in Refs.~\cite{PhysRevB.75.155111,PhysRevLett.109.017202,PhysRevX.5.041047,PhysRevLett.110.260601,PhysRevLett.117.160601}.

	 	 \begin{figure*}[t]     		
	 	 	\includegraphics[width=0.9\textwidth]{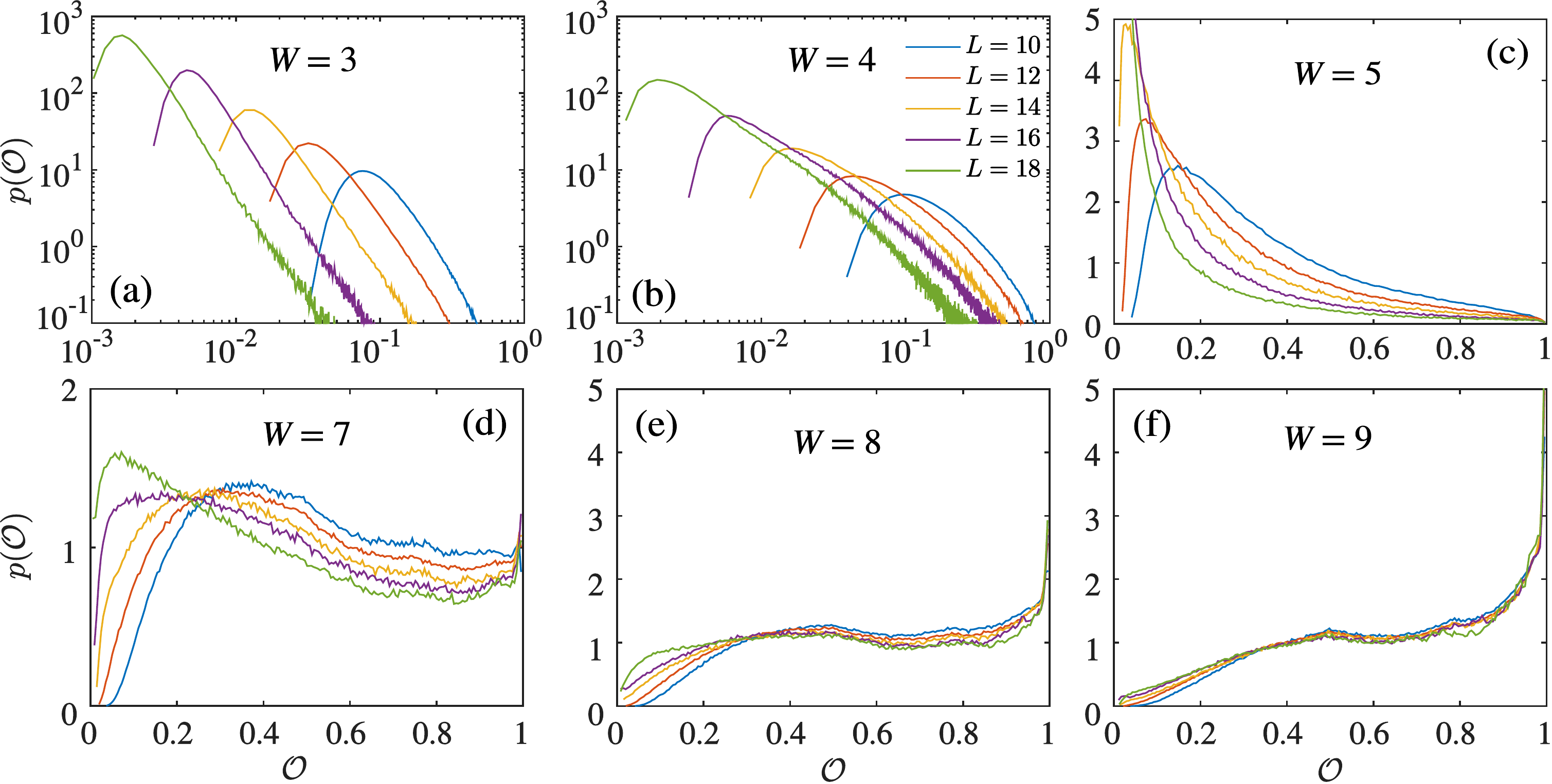}
	 	 	\caption{The distribution of eigenstate overlaps across a MBL transition for different system sizes with $J_z =1$.
	 	 		(a)-(c) For weak disorder ($W=3,4,5$), the system is in the ergodic phase. When increasing the system size, the curves of 
	 	 		$p(\mathcal{O})$ have an exponentially top-left shift.
	 	 		(d) In the case of a critical point ($W=7$), the top-left shifts almost vanish.
	 	 		(e),(f) For strong disorder ($W=8,9$), the systems are in the MBL phase.
	 	 		In this case, the curves of $p(\mathcal{O})$ for different system sizes almost coincide.
	 	 	}
	 	 	\label{fig_2}
	 	 \end{figure*}

	 By a Jordan-Wigner transformation, Hamiltonian (\ref{H}) can be mapped to a local spinless fermionic 
	 system with a nearest-neighbor hopping strength $J$ and 
	 density-density interaction strength $J_z$.
	 Below, for convenience, we set $J=1$.
	 When $J_z=0$, it is a free system with a disorder potential, and in this case, 
	 all the single-particle eigenstates are localized  when $W\neq0$, i.e., it is in the AL phase.
	 When $J_z \neq 0$, this becomes an interacting model, where the single-particle description may fail.
	 For different disorder strengths, this system can be divided into three regimes:
	 At weak disorder ($0<W<W_1$), it belongs to the ergodic phase with all of the eigenstates ergodic, at strong disorder ($W>W_2$), it belongs to the MBL phase with all of the eigenstates localized,
	 and at intermediate disorder strength ($W_1<W<W_2$), the system has many-body mobility edges and is in the Griffiths phase~\cite{PhysRevLett.114.160401,PhysRevB.94.144201,PhysRevB.96.104201}.
	 Especially, when $J_z=1$, i.e., the Hamiltonian is Heisenberg coupled, the critical disorder strengths are $W_1 \approx 4$ and $W_2 \approx 7$, respectively~\cite{PhysRevB.82.174411,PhysRevLett.114.160401,PhysRevX.5.041047,PhysRevB.94.144201,PhysRevB.96.104201}.

	 		\begin{table}[b]
	 			\centering
	 			\scriptsize
	 			\renewcommand\arraystretch{1.6}
	 			\resizebox{0.5\textwidth}{!}{
	 				\begin{tabular}{c |c c c c c }
	 					\hline
	 					\hline
	 					
	 					&$L=10$  &$L=12$  &$L=14$  &$L=16$  &$L=18$\\
	 					\hline
	 					$J_z=1,W<6$  &$30000$  &$10000$  &$1000$  &$300$  &$100$\\
	 					$J_z=1,W\geq6$  &$60000$  &$20000$  &$2000$  &$600$  &$200$\\
	 					$J_z=0$  &$60000$  &$20000$  &$2000$  &$600$  &$200$\\
	 					\hline
	 					\hline
	 					
	 				\end{tabular}
	 				
	 			}
	 			\caption{The numbers of disorder averaging for different parameters of Hamiltonian (\ref{H}).
	 			}
	 			\label{tab1}	
	 		\end{table}

    \subsection{Methods}
	 Here, we focus on the sensitivity of many-body wave functions with respect to the boundary conditions.
	 We choose a periodic boundary condition, i.e., $\hat{\sigma}_{L+1}=\hat{\sigma}_{1}$, 
	 and antiperiodic boundary condition, i.e., $\hat{\sigma}_{L+1}=-\hat{\sigma}_{1}$.
	 To quantify the robustness of many-body wave functions, 
	 we define the overlaps between the two corresponding eigenstates with different boundary conditions,
	 \begin{align} \label{On} 
	 	 \mathcal{O}_n = |\langle\psi_n^{\text{p}}|\psi_n^{\text{ap}}\rangle|^2,
	 \end{align}
	 where $|\psi_n^{\text{p}}\rangle$ and $|\psi_n^{\text{ap}}\rangle$ are the eigenstates of $\hat H$
	 with periodic and antiperiodic boundary conditions, respectively.
	 Generally, the change in the boundary condition can be regarded as applying a local perturbation.
	 Therefore, according to the perturbation theory, $\sqrt{\mathcal{O}_n}|\psi_n^{\text{p}}\rangle$ is the zero-order contribution (the phase is neglected) of $|\psi_n^{\text{ap}}\rangle$, and vice versa.
	 If the eigenstates of the system are robust to the boundary conditions, 
	 then the eigenstate overlap $\mathcal{O}$ will be large, i.e., $\mathcal{O}$ can reflect the sensitivity of eigenstates with respect to the boundary conditions.

	 In addition, we can define the distribution density of $\mathcal{O}_n$ as
	 \begin{align} \label{Po} 
	 p(\mathcal{O}) \equiv \frac{1}{\mathcal{N}}\sum_{n=1}^{\mathcal{N}}\delta(\mathcal{O}_n-\mathcal{O}),
	 \end{align}
	 where $\mathcal{N}$ is the number of eigenstate pairs.
	 The mean of $\mathcal{O}_n$ can also be obtained as 
	 \begin{equation} \label{Pm} 
	 \bar{\mathcal{O}} \equiv \frac{1}{\mathcal{N}}\sum_{n=1}^{\mathcal{N}}\mathcal{O}_n = \int_0^1 d\mathcal{O}\ \mathcal{O} p(\mathcal{O}).
	 \end{equation}
	 Thus, we can use $\bar{\mathcal{O}}$ to analyze the robustness of the eigenstates, quantitatively.

	 We use exact diagonalization to extract the eigenstates of the random-field XXZ Hamiltonian (\ref{H})
	 with periodic and antiperiodic boundary conditions, respectively.
	 For each diagonalization, two boundary conditions should satisfy the same disorder configuration,
	 and due to the spin $U(1)$ symmetry, we only consider the eigenstates with half filling.
	  We sort the many-body eigenstates with the corresponding eigenenergies from small to large, and the $m$th eigenstate for the antiperiodic boundary condition is considered as the corresponding state of the $n$-th eigenstate with a periodic boundary condition.
	 For a quantum many-body system, the eigenenergies are very dense with the differences in the energy level $\Delta E_n\sim e^{-L}$. 
     Thus,  $|E_{m}^\text{ap}-E_{n}^\text{p}|\sim 1\gg \Delta E_{n}^\text{p},\Delta E_{m}^\text{ap}$, so that it is possible that $m\neq n$. Therefore, it is hard to judge which two eigenstates with different boundary conditions are related.
	  Here, to obtain $\mathcal{O}_n$, for each eigenstate $|\psi_n^{\text{p}}\rangle$, 
	 we need to calculate the overlaps with all eigenstates of $\hat H$ with an antiperiodic boundary condition.
	 We regard the maximum among these overlaps as $\mathcal{O}_n$, i.e.,
	 \begin{equation} \label{On2} 
	  \mathcal{O}_n = \text{max} \{ |\langle\psi_n^{\text{p}}|\psi_1^{\text{ap}}\rangle|^2,|\langle\psi_n^{\text{p}}|\psi_2^{\text{ap}}\rangle|^2, ... ,|\langle\psi_n^{\text{p}}|\psi_{\mathcal{D}}^{\text{ap}}\rangle|^2 \},\\
	 \end{equation}
	 where $\mathcal{D} = \binom{L}{L/2}$ is the dimension of half-filling Hilbert space. Thus, $\mathcal{O}_n$ can be considered as the maximum entries in each row of the \textit{overlap matrix} defined in Ref.~\cite{PhysRevE.86.046202}.
	 Intuitively, this definition can indeed represent the robustness of the many-body eigenstates.
	 Comparing with the original definition of $\mathcal{O}_n$, i.e., Eq.~(\ref{On}), this method seems to result in an increase in the value of $\mathcal{O}_n$.
	 However, in the following discussion, we are mainly concerned about the distribution and scaling of $\mathcal{O}_n$ rather than the explicit values of $\mathcal{O}_n$, which can represent the robustness and independence of the many-body eigenstates.
     Therefore, this method is somehow reasonable in terms of the topics in this work.

	 To avoid any possible influence, such as many-body mobility edges, 
	 we focus on the middle one-eighth of the full eigenstates.
	 In this case, the Griffiths phase at intermediate disorder strength can be neglected, 
	 and the MBL transition occurs at $W\approx7$ with $J_z = 1$~\cite{PhysRevB.75.155111}.
	 The phase diagrams of the Hamiltonian (\ref{H}) are presented in Fig.~\ref{fig_1}(b).
	 The numbers of disorder averaging for different sizes and disorder strengths are presented in Tabble.~\ref{tab1}.

\begin{figure}[t]     	
	\includegraphics[width=0.35\textwidth]{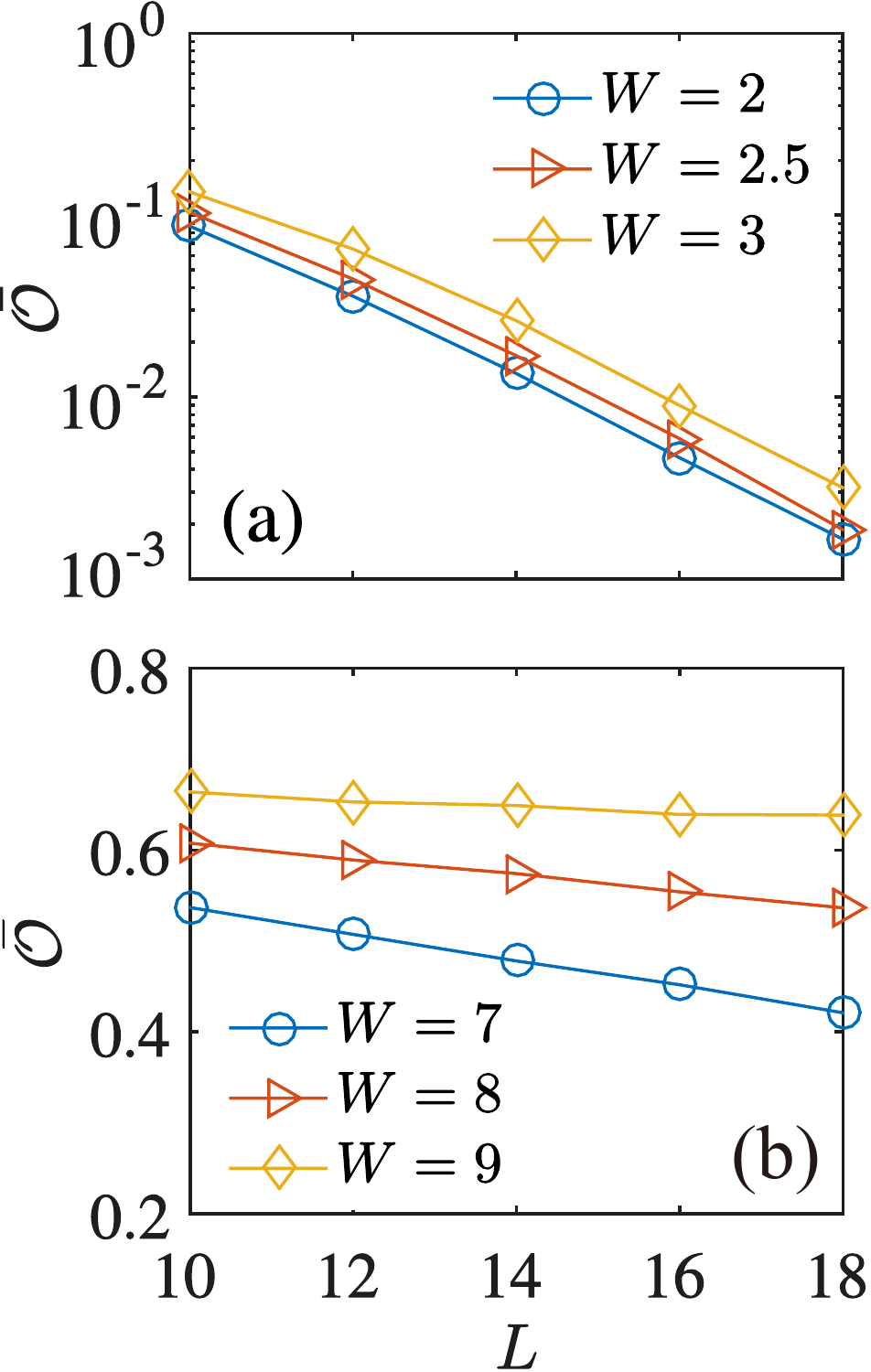}	
	\caption{The scaling of $\bar{\mathcal{O}}$ for different phases.
		(a) In the ergodic phase, $\bar{\mathcal{O}}$ is exponential decay with the increase of systems sizes.
		(b) In the MBL phase, $\bar{\mathcal{O}}$ almost keeps invariant, when increasing the system size.}
	\label{fig_3}
\end{figure}

\begin{figure}[t]     		
	\includegraphics[width=0.35\textwidth]{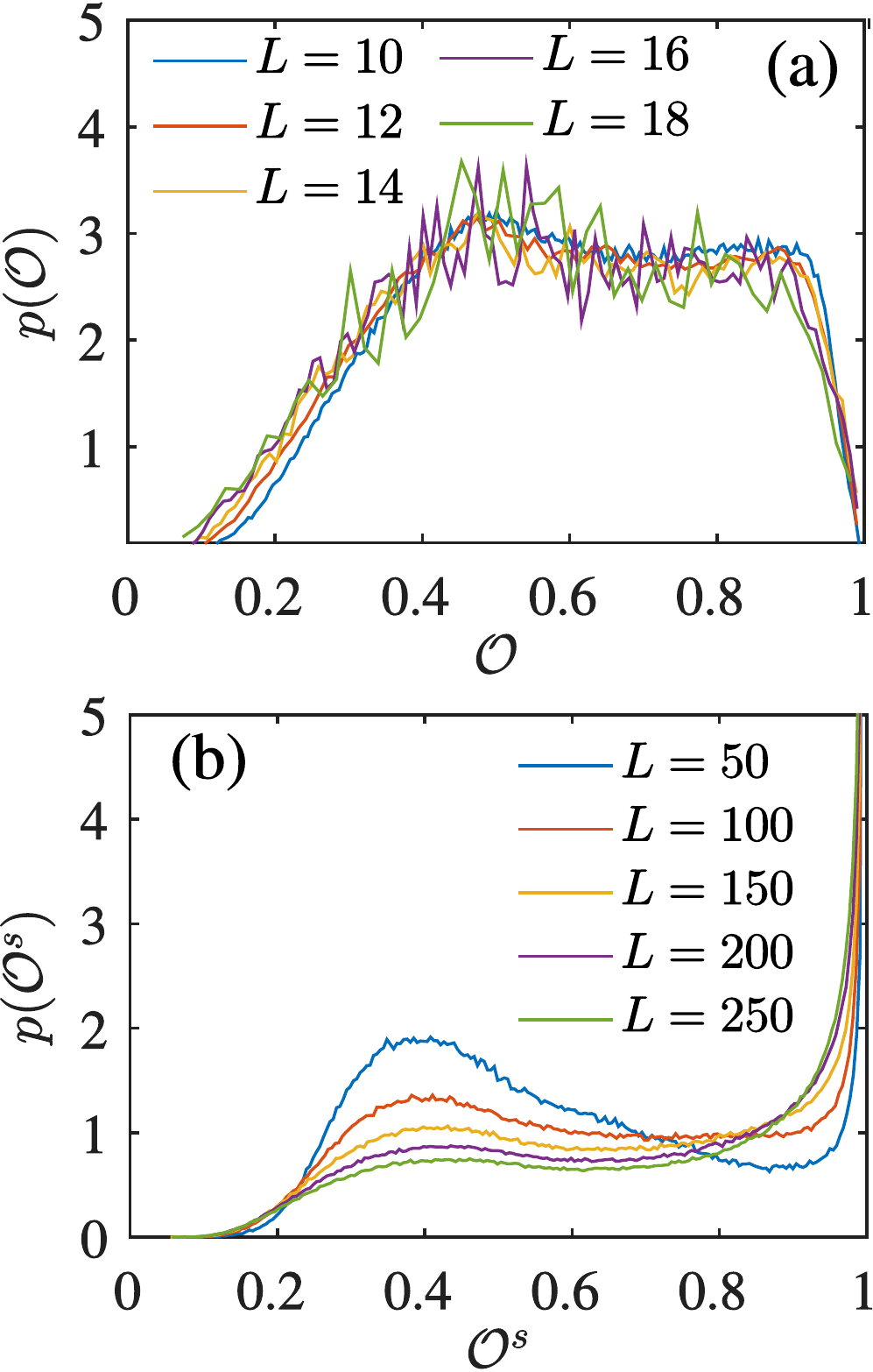}
	\caption{(a) The distribution of the overlaps of many-body wave functions with respect to 
		the periodic and anti-periodic boundary conditions for $H$ with $J_z =0$. Here, the system is in AL phase.
		(b)  The distribution of the overlaps of single-particle eigenstates with respect to 
		the periodic and anti-periodic boundary conditions for $\hat H_{\text{F}}$ with $W = 0.5$.
		Here, all of the single-particle eigenstates for each system are considered.		
	}
	\label{fig_4}
\end{figure}

	 \section{Numerical results}\label{Numerical results}
	 First, we consider an interacting system with $J_z=1$, i.e., the Hamiltonian is Heisenberg coupled. 
	 As mentioned, in this case, the localization-delocalization phase transition point is at $W_2 \approx 7$.
	 In Fig.~\ref{fig_2}, we show the distribution density of eigenstate overlaps $p(\mathcal{O})$ with different system sizes in both delocalized and localized phases. 
	 With an increase of disorder strength, the curves of $p(\mathcal{O})$ have a right shift, which indicates that the robustness of the eigenstates becomes stronger when increasing the disorder strength. 
	 Additionally, when increasing the system size, $p(\mathcal{O})$ exhibits a top-left shift in the delocalized phase [see Figs.~\ref{fig_2}(a)-(c)].
	 In Fig.~\ref{fig_2}(d), we can find that this shift almost vanishes near the critical point.
	 In the MBL phase,  $p(\mathcal{O})$ is almost size independent, since the curves of $p(\mathcal{O})$ for different system sizes nearly coincide, see Figs.~\ref{fig_2}(e,f).

\begin{figure*}[t]     		
	\includegraphics[width=0.9\textwidth]{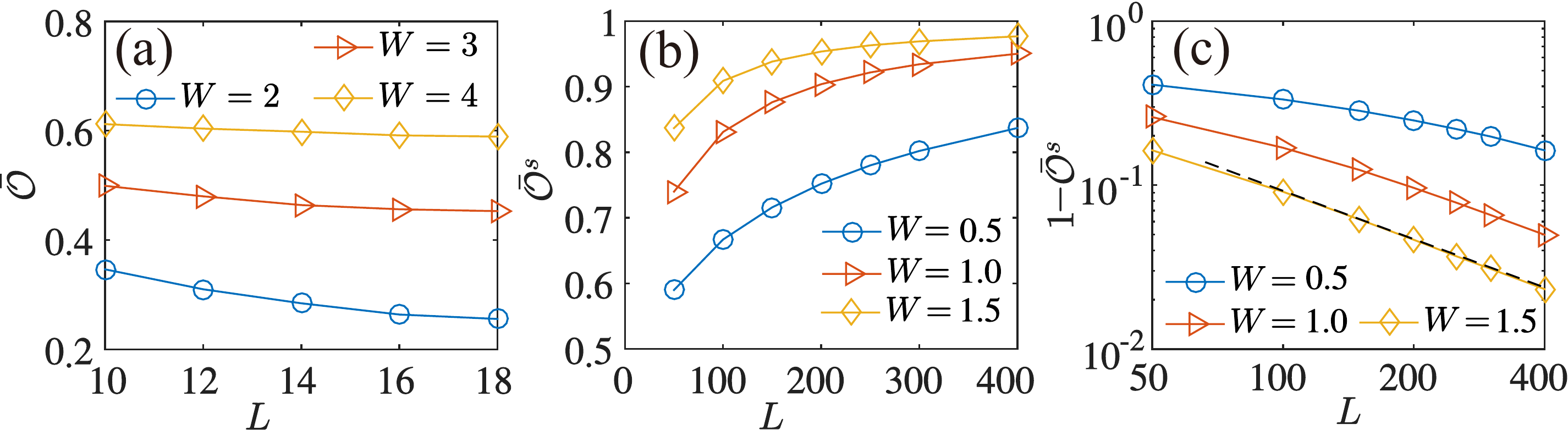}
	\caption{
		The scaling of (a) $\bar{\mathcal{O}}$, (b) $\bar{\mathcal{O}}^s$, and (c) $1-\bar{\mathcal{O}}^s$ for AL systems, respectively. For (c), the black dashed line is a linear fitting, which represents $1-\bar{\mathcal{O}}^s\propto L^{-1}$.		
	}
	\label{fig_5}
\end{figure*}

	 To extract the scaling of the eigenstate overlaps, we calculate the means of $\mathcal{O}_n$, 
	 i.e., $\bar{\mathcal{O}}$.
	 In Fig.~\ref{fig_3}, we carry outa finite-size scaling analysis of $\bar{\mathcal{O}}$ between the ergodic  and MBL phase.
	 For the ergodic systems, see Fig.~\ref{fig_3}(a), we can find that $\bar{\mathcal{O}}$ is exponentially dependent on the system size
	 \begin{equation}  
	 	 \bar{\mathcal{O}}_{\text{Er}} \propto e^{-\alpha L}.
	 \end{equation}
	 This indicates that the many-body wave functions of ergodic systems are very sensitive to the boundary conditions.	 
	 Thus, in the thermodynamic limit, the eigenstates of the ergodic systems will become completely different when applying small local perturbations, and this can be considered as a butterfly effect of eigenstates in ergodic systems.
	 This exponential scaling of $\bar{\mathcal{O}}$ resembles an Anderson orthogonality catastrophe~\cite{PhysRevLett.18.1049}, which  involves the ground state of quantum many-body systems.

	 For the MBL phase, according to Fig.~\ref{fig_3}(b), $\bar{\mathcal{O}}$ is almost size independent 
	 \begin{align} 
	 \bar{\mathcal{O}}_{\text{MBL}} \propto O(1).
	 \end{align}
	 Therefore, the many-body wave functions in MBL systems are robust with respect to the boundary conditions.

	 Now we take up the non-interacting systems, where $J_z=0$.
	 In this case, the systems are in the AL phase for arbitrary weak disorder.
	 In Fig.~\ref{fig_4}(a), we show the distribution of the  eigenstate overlaps of this system.
	 Comparing with Figs.~\ref{fig_2}(e,f), we find that the curves of $p(\mathcal{O})$ in the AL phase are distinct from the MBL phase.

	 To further uncover the properties of many-particle eigenstates in an  AL system, we use a single-particle representation to study this system.
	 The Hamiltonian here can be mapped to a free spinless fermion system with  the Hamiltonian
	 \begin{align} 
	  \hat H_{\text{F}} = \sum_{i}^{L} (\hat{c}^\dagger_{i}\hat{c}_{i+1}+\hat{c}^\dagger_{i+1}\hat{c}_{i})
	 	 +\sum_{i=1}^Lh_{i}\hat{c}^\dagger_{i}\hat{c}_{i},
	 \end{align}
	 where $\hat{c}^\dagger_{i}$ ($\hat{c}_{i}$) is the creation (annihilation) operator of fermions.
	 In Fig.~\ref{fig_4}(b), we present the corresponding distribution of single-particle wave functions $\mathcal{O}^s$.
	 We can find that the curves of $p(\mathcal{O}^s)$ of $\hat H_{F}$ are similar to the curves of $p(\mathcal{O})$ in MBL systems shown in Figs.~\ref{fig_2}(e,f).

	 In addition, we  calculate the the means of the overlaps for both many-body wave functions and single-particle wave functions in AL systems, see Fig.~\ref{fig_5}(a,b). We find they are both almost size-independent
	 \begin{align} 
	 	\bar{\mathcal{O}}_{\text{AL}} ,\bar{\mathcal{O}}^{s}_{\text{AL}} \propto O(1),
	 \end{align}
	 showing the robustness of both many-body and single-particle eigenstates to the boundary conditions, which are identified with the MBL systems.

	 \section{Phenomenological discussion}\label{Phenomenological discussion}
	 In the last section, we have presented the main numerical results.
	 Here, based on these numerical results, we give some phenomenological interpretations.
	 We demonstrate that the fragility of many-body eigenstates in ergodic systems 
	 is consistent with the random matrix theory, 
	 and the independence of many-body eigenstates in the MBL systems is consistent with the existence of a large number of quasilocal integrals of motion.

	 \subsection{Delocalized Systems}
	 We know that  ergodic systems, as quantum chaos, can be described by the random matrix theory.
	 According to the random matrix theory, the spectrum of the ergodic systems satisfies the Wigner-Dyson distribution, 
	 and the eigenstates are very sensitive to small perturbations~\cite{RevModPhys.53.385, Luca2016}.
	 This sensitivity of many-body wave functions in ergodic systems can indeed be represented by our results shown in the last section.

     Now we analyze the scaling of $\bar{\mathcal{O}}_{\text{Er}}$ by means of perturbation theory and random matrix theory, qualitatively.
	 The difference between the periodic and anti-periodic boundary conditions is a local perturbation $V$.
     According to the first-order perturbation theory, we have 
     \begin{align} 
     |\psi_n^{\text{ap}}\rangle = |\psi_n^{\text{p}}\rangle +
      \sum_{m\ne n}C_{mn}|\psi_m^{\text{p}}\rangle,
     \end{align}
	 where $C_{mn}\equiv V_{mn}/(E_n^{\text{p}}-E_m^{\text{p}})$, $V_{mn} \equiv \langle\psi_m^{\text{p}}|V|\psi_n^{\text{p}}\rangle$
	 and $E_n^{\text{p}}$ is the corresponding eigenenergy of $|\psi_n^{\text{p}}\rangle$.
	 In the ergodic phase, using Srednicki's ansatz~\cite{Srednicki_1999}, 
	 we have the off-diagonal matrix elements of the local operator
	 \begin{align} 
	   V_{mn} = e^{-S(E,L)/2}f(E_m,E_n)R_{mn},
	 \end{align}
	 where $S(E,L)$ is the statistical entropy at energy $E=(E_n+E_m)/2$, 
	 $R_{mn}$ is a random matrix with order one, 
	 and $f$ is a smooth function. Generally, $S(E,L)=\varepsilon\ln\mathcal{D}$ with $\varepsilon\leq1$.
	 Thus, $|C_{mn}|^2\propto\mathcal{D}^{-\varepsilon}$, and $\sum_{m=1}^{\mathcal{D}}|C_{mn}|^2\propto\mathcal{D}^{1-\varepsilon}$.
	 Therefore, for the ergodic systems, $\mathcal{O}_n\propto\mathcal{D}^{\varepsilon-1}$, which indicates $\bar{\mathcal{O}}_{\text{Er}}$ decays exponentially with an increase in the system size.

	 \subsection{Localized Systems}
	 For AL systems, the single-particle eigenstates satisfy $\bar{\mathcal{O}}^{s}_{\text{AL}} \propto 1-c/L$,
	 see Fig.~\ref{fig_5}(c).
	 Thus, for many-body eigenstates, which can be written as Slater determinants, 
	 we have $\bar{\mathcal{O}}_{\text{AL}} \sim (1-c/L)^L\sim O(1)$.

	 Comparing MBL with AL systems, we find that the behaviors of the half-filling eigenstates in the MBL phase
	 are more similar to the single-particle rather than the many-body eigenstates of the AL phase.
	 In fact, this is consistent with the existence of quasilocal integrals of motion in MBL systems.
	 In the AL phase, the single-particle eigenstates are local conservation modes, 
	 and different modes are decoupled.
	 Thus, the many-body eigenstates, as the Slater determinants, are not independent of each other.
	 In contrast, for MBL systems, there exist a large number of quasilocal integrals of motion, 
	 which are the many-body modes and independent of each other.
	 Therefore, it is reasonable that the many-body eigenstates for the MBL systems behave more as the single-particle one in AL systems.
	 
	 To further illustrate our results, in Appendix \ref{appA}, we study the transverse field Ising model with disorder at longitudinal field. The numerical results of this model are consistent with the above discussions,
	 which indicates the universality of our results for different models.
	 
\begin{figure*}[t]     		
	\includegraphics[width=0.6\textwidth]{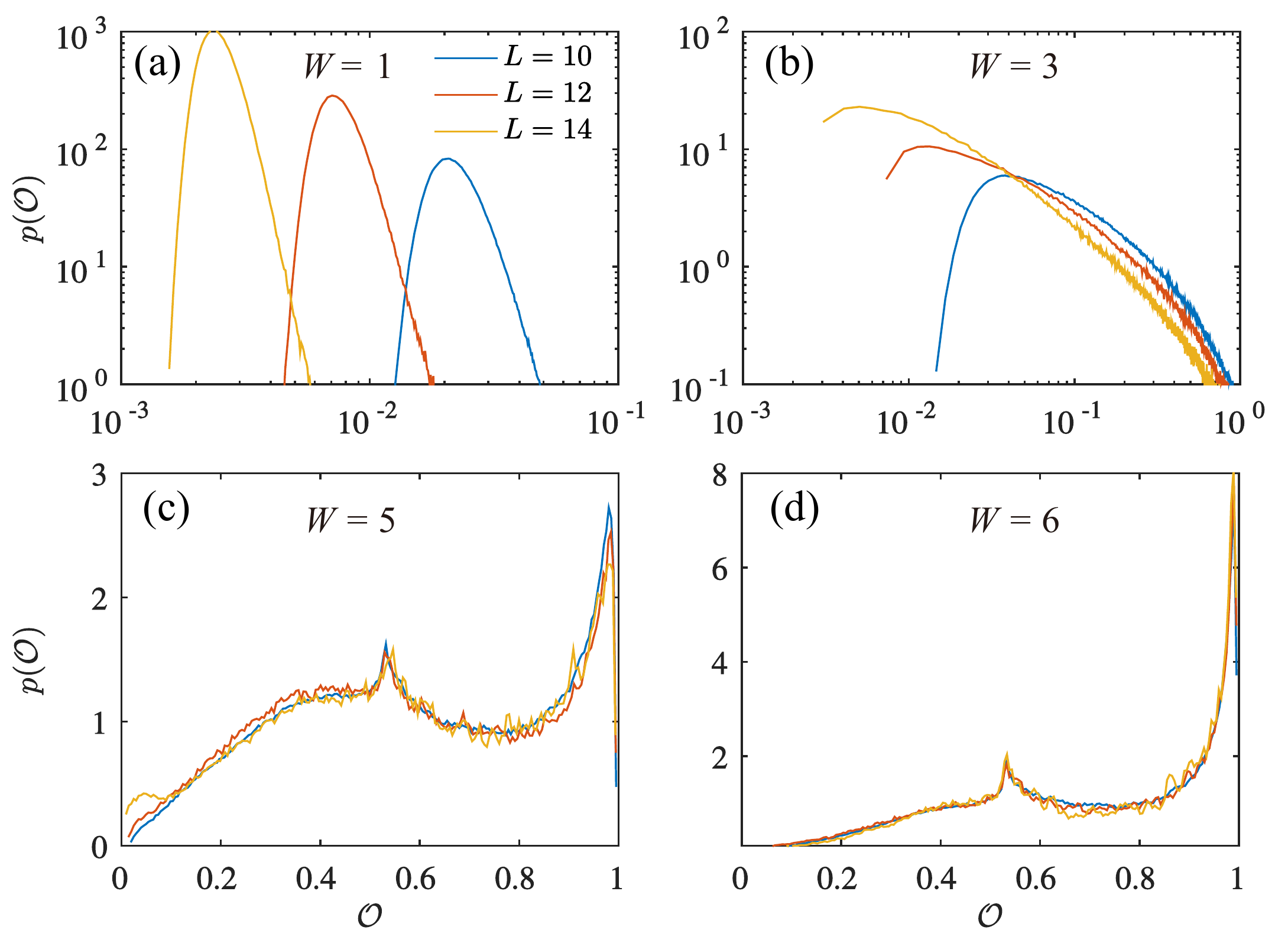}
	\caption{The distribution of eigenstate overlap in the delocalized and localized systems of Hamiltonian (\ref{Hmi}) for different system sizes.
		(a,b) For weak disorder ($W=1,3$), the system is in the ergodic phase. When increasing the system size, the curves of 
		$p(\mathcal{O})$ shift exponentially to top-left.
		In addition, the curves of $p(\mathcal{O})$ here are similar to the case of the random-field XXZ model 
		showing in Figs.~\ref{fig_2}(a-c).
		(c,d) For strong disorder ($W=5,6$), the systems are in the MBL phase.
		In this case, the curves of $p(\mathcal{O})$ for different system sizes nearly coincide,
		and they are also consistent with the MBL phases of the random-field XXZ model.
		Here, the numbers of disorder averaging are 10000, 1000 and 300, when $L =10,12,14$, respectively. 
		In addition, only the middle one-eighth of the full eigenstates are considered.}
	\label{fig_a1}
\end{figure*}

	 \section{Conclusion}\label{Conclusion}
	 In summary, we have studied the sensitivities of the eigenstates to the boundary conditions between  localized and delocalized systems.
	 By calculating the overlaps of the corresponding eigenstates between periodic and anti-periodic boundary conditions, we find that the eigenstates are robust to the boundary conditions
	 in localized phases, while they are fragile in delocalized phases.
	 Furthermore, the many-body eigenstates in MBL systems have similar behaviors to the single-particle eigenstates in AL system, and this is consistence with the existence of a large number of quasilocal integrals of motion in the MBL phase. 
	 Our results provide another viewpoint to explore the MBL, i.e., directly from the many-body eigenstates.

	 Generally, there are many methods to diagnose whether a system is localized or delocalized, for instance, the level statistic and the real-time dynamics.
	 However, there are few ways to diagnose the MBL and AL, and the sole method may be to see whether the spreading of the entanglement entropy is unbounded.
	 In this work, on the one hand, according to the scaling of the many-body eigenstates overlaps $\mathcal{O}_n$, we can diagnose whether the system is localized.
	 On the other hand, according to the distribution of $\mathcal{O}_n$, we can distinguish the MBL and AL.
     In addition, our results can also be considered as a signature of the existence of many quasilocal integrals of motions in the MBL systems.

	 Finally, there remains an open problem as to whether our results are related to the nontrivial dynamics of MBL systems, such as the logarithmic spreading of entanglement entropy.

\begin{acknowledgements}
	This work was supported by National Key R \& D Program of China (Grant Nos. 2016YFA0302104 and 2016YFA0300600), National
	Natural Science Foundation of China (Grant Nos. 11774406 and 11934018), Strategic
	Priority Research Program of Chinese Academy of Sciences (Grant No. XDB28000000),
	and Beijing Academy of Quantum Information Science (Grant No. Y18G07).
\end{acknowledgements}

\appendix

\section{Mixed Transverse Field Ising Model}\label{appA}	
Here, we study another spin model, i.e., the transverse field Ising model with disorder at the longitudinal field, to further illustrate our results.
The corresponding Hamiltonian reads
\begin{align} \label{Hmi}
\hat  H_{\text{MI}} = \sum_{i}^{L}\hat{\sigma}^z_{i} \hat{\sigma}^z_{i+1}+h_{x}\sum_{i=1}^L\hat{\sigma}^x_{i}
 +\sum_{i=1}^Lh_{z,i}\hat{\sigma}^z_{i},
\end{align}
where the $z$-directional random-field $h_{z,i} \in [-W+\bar{h}_z,W+\bar{h}_z]$ satisfies a uniform distribution.
Here, we choose the parameters $h_x = 1.05$ and $\bar{h}_z=0.5$.
According to Ref.~\cite{PhysRevLett.123.165902}, the critical point of the localization-delocalization phase transition is $W = 4.2$.
At the weak disorder regime ($W<4.2$), the system is in the ergodic phase.
At strong disorder regime ($W>4.2$), it is in the MBL phase.

Following the numerical method mentioned in the main text, we calculate the eigenstates overlaps  $\mathcal{O}$ with respect to the periodic and antiperiodic boundary conditions.
In Fig.~\ref{fig_a1}, we present the distributions of the eigenstates overlaps, i.e., $p(\mathcal{O})$.
In the ergodic phase, we can find that $p(\mathcal{O})$ has a exponential top-left shift with the increase of system size,
see Figs.~\ref{fig_a1}(a,b).
According to Figs.~\ref{fig_a1}(c,d), in the MBL case,  $p(\mathcal{O})$ are almost size-independent, and the curves of $p(\mathcal{O})$ are also similar to the single-particle cases of AL systems shown in Fig.~\ref{fig_4}(b).
Therefore, comparing with Fig.~\ref{fig_2}, we can find that the numerical results of Hamiltonian (\ref{Hmi}) are closely consistent with the cases of the random-field XXZ chain.

\end{document}